\documentclass[journal,twocolumn,10pt]{IEEEtran}
\normalsize
\usepackage[noadjust]{cite}
\usepackage{filecontents}
\usepackage{color,colortbl}
\usepackage[pdftex]{graphicx}
\usepackage{verbatim} 
\usepackage{epstopdf}
\usepackage[cmex10]{amsmath}
\usepackage[tight,footnotesize]{subfigure}
\usepackage{pifont}
\usepackage{algorithm}
\usepackage{algpascal}
\usepackage{algpseudocode}
\usepackage{multirow}
\usepackage{epstopdf}
\usepackage[colorlinks=true,pdfstartview=FitV,linkcolor=black,citecolor=black, urlcolor=blue,plainpages=false]{hyperref}
\usepackage{hyperref}
\definecolor{Gray}{gray}{0.99}
\newcommand{\E}{{\rm I\kern-.3em E}}

\begin{document}
\title{Digital Predistortion for Hybrid MIMO Transmitters}
\author{Mahmoud~Abdelaziz,~\IEEEmembership{Member,~IEEE,}
        Lauri~Anttila,~\IEEEmembership{Member,~IEEE,}
				Alberto~Brihuega,~\IEEEmembership{Student Member,~IEEE,}
				Fredrik~Tufvesson,~\IEEEmembership{Fellow,~IEEE,}
        Mikko~Valkama,~\IEEEmembership{Senior~Member,~IEEE}
\thanks{Mahmoud~Abdelaziz, Lauri~Anttila, Alberto~Brihuega, and Mikko~Valkama are with the Laboratory of Electronics and Communications Engineering, Tampere University of Technology, Tampere, Finland.} 
\thanks{Fredrik~Tufvesson is with the Department of Electrical and Information Technology, Lund University, Lund, Sweden.} 
\thanks{This work was supported by Tekes, Nokia Bell Labs, Huawei Technologies Finland, TDK-EPCOS, Pulse Finland and Sasken Finland under the 5G TRx project, and by the Academy of Finland under the projects 288670 ``Massive MIMO: Advanced Antennas, Systems and Signal Processing at mm-Waves", 284694 ``Fundamentals of Ultra Dense 5G Networks with Application to Machine Type Communication", and 301820 ``Competitive Funding to Strengthen University Research Profiles".}
}
\maketitle
\begin{abstract}
This article investigates digital predistortion (DPD) linearization of hybrid beamforming large-scale antenna transmitters. We propose a novel DPD processing and learning technique for an antenna sub-array, which utilizes a combined signal of the individual power amplifier (PA) outputs in conjunction with a decorrelation-based learning rule. In effect, the proposed approach results in minimizing the nonlinear distortions in the direction of the intended receiver. This feature is highly desirable, since emissions in other directions are naturally weak due to beamforming. The proposed \textcolor{black}{parameter learning} technique requires only a single observation receiver, and therefore supports simple hardware implementation. It is also shown to clearly outperform the current state-of-the-art technique which utilizes only a single PA for learning. Analysis of the feedback network amplitude and phase imbalances reveals that the technique is robust even to high levels of such imbalances. {\color{black}Finally, we also show that the array system out-of-band emissions are well-behaving in all spatial directions, and essentially below those of the corresponding single-antenna transmitter, due to the combined effects of the DPD and beamforming.}
\end{abstract}


\IEEEpeerreviewmaketitle

\section{Introduction}

\begin{figure}[t!]
\centering
\subfigure[Digital MIMO transmitter architecture with per antenna/PA digital predistortion.]{
\includegraphics[width=0.9\linewidth]{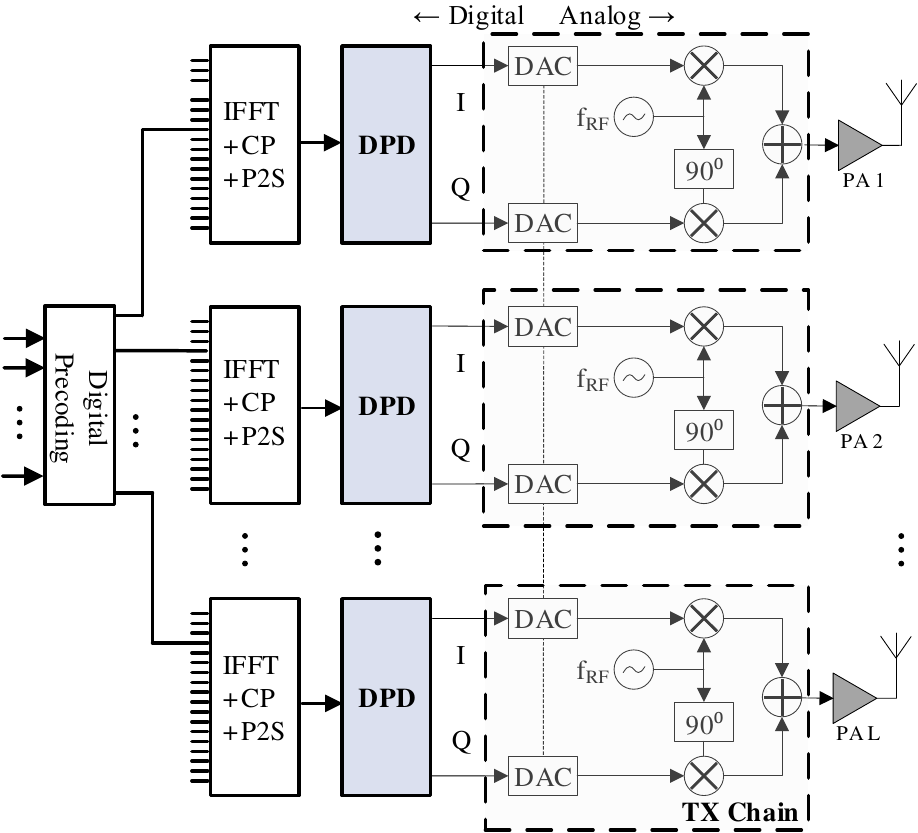}
\label{fig:TX_Digital_MIMO}
}
\subfigure[Hybrid MIMO transmitter architecture with per sub-array digital predistortion.]{
\includegraphics[width=1\linewidth]{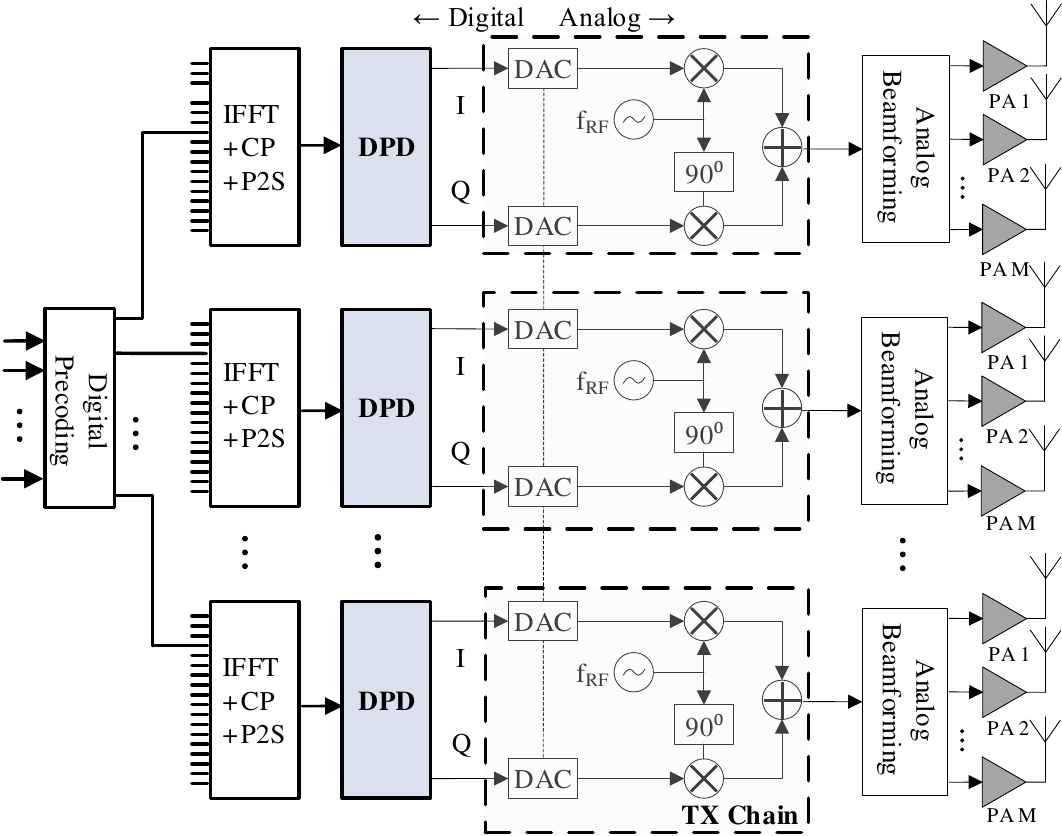}
\label{fig:TX_Hybrid_MIMO}
}
\caption{Digital versus Hybrid MIMO transmitter architectures. Thick lines correspond to complex I/Q processing.}
\label{fig:TX_arch}
\end{figure}

\IEEEPARstart{L}{arge}-scale antenna systems are expected to be one of the key enablers of enhanced spectral and energy efficiency in future wireless communication systems \cite{Intro_1,Intro_2}. Utilizing fully-digital beamforming at the transmitter, as most works assume, would mean that each antenna should have a dedicated transmit chain, as depicted in Fig. \ref{fig:TX_Digital_MIMO}. To relieve the large hardware costs of such implementations, there has been increasing interest on splitting the beamforming operation between digital and analog domains \cite{Intro_3}. One possible implementation of such a \emph{hybrid beamforming} transmitter is shown in Fig. \ref{fig:TX_Hybrid_MIMO}. The overall transmitter contains antenna subsystems of $M$ antennas, which are connected to a single RF transmitter chain via an analog beamforming unit.

The power efficiency of the transmitters is very important in future massive antenna arrays with several hundreds of megahertz instantaneous transmit bandwidth, since the consumed energy per bit is to be kept constant or preferably even lowered compared to 4G systems \cite{Intro_1,Intro_2}. The power efficiency of the PAs, which are the most power hungry components in the transmitter (independent of whether fully-digital or hybrid beamforming is used), therefore needs to be high. Thus, low-cost, small-size and highly energy-efficient, and therefore highly nonlinear PAs operating close to saturation, are expected to be adopted.

Some recent studies have investigated the impact of PA nonlinearities on massive MIMO transmitters \cite{Intro_4,Intro_5,Intro_6,Intro_7,Intro_8,Intro_9,OOB_Mollen}. These studies show that the spectral efficiency and the energy efficiency, both of which are fundamental objectives of massive MIMO, are compromised. In \cite{Intro_5}, the out-of-band radiation due to PA nonlinearity was analyzed in both single antenna and massive MIMO transmitter scenarios, assuming a memoryless polynomial model for each PA unit. It was shown that the adjacent channel leakage ratio (ACLR) due to PA nonlinearity in the massive MIMO scenario is, on average, equal to the single antenna scenario when transmitting with the same total sum-power. This implies that when a highly nonlinear PA is used per RF chain, as mentioned earlier, significant out-of-band distortion can occur in massive MIMO transmitters that can easily interfere with neighboring channel transmissions and/or violate the spurious emission limits, as also demonstrated in \cite{Intro_6}.

In terms of the impact of hardware impairments on the transmitted signal quality, it was shown in \cite{Intro_6} that the error vector magnitude (EVM) degradation due to PA nonlinearity can compromise the spectral efficiency of the massive MIMO base station. In \cite{Intro_6}, at least 6 dB backoff was shown to be required in order to reach the maximum targeted data rate. Moreover, in \cite{Intro_7}, the authors demonstrated that when practical PA models are used in a massive MIMO base station, the signal to interference and noise ratio (SINR) at the user receiver could be significantly degraded.

\begin{figure}[t!]
\centering
\includegraphics[width=1\linewidth]{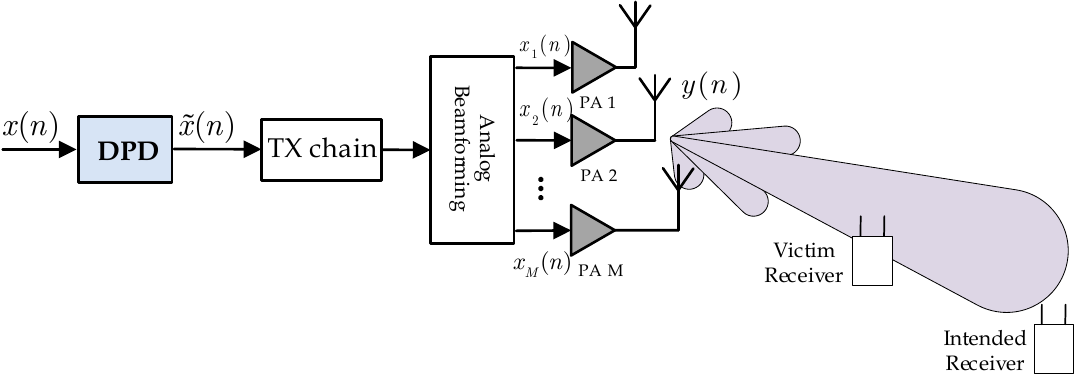}
\caption[]{Block diagram of a single sub-array in a hybrid MIMO transmitter. The effective signal radiated towards the intended RX direction, $y(n)$, is the superposition of the individual antenna outputs when assuming ideal beamforming and free-space LoS conditions. The worst-case victim RX, in terms of OOB radiation, lies also in the direction of the main beam.}
\label{fig:TX_subarray_main_path}
\end{figure}

In \cite{OOB_Mollen}, a more detailed study was conducted regarding the out-of-band radiation in massive MIMO transmitters when the PA nonlinearity is considered. It was shown in \cite{OOB_Mollen} that when assuming a single user per array, and free-space line-of-sight (LoS) propagation with ideal beamforming, the most harmful emissions are in the same direction as the main beam. It was also shown that under this assumption, the in-band and out-of-band unwanted emissions due to the nonlinear PAs are identical to the single antenna case in the direction of the main beam towards the intended RX. Thus, the worst case scenario will occur when a victim user lies in the same direction as the intended user, as shown in Fig. \ref{fig:TX_subarray_main_path}.

In general, applying backoff to overcome the PA distortion is not an attractive solution since it requires using larger PAs operating in the linear region, assuming a given transmit sum-power requirement. As a result, the cost and size of each RF chain would increase and the energy efficiency would decrease, which directly translates to increased running costs in terms of power supply and cooling. Thus a more intriguing solution is to use smaller PAs that operate more efficiently close(r) to saturation, while using a low complexity linearization method to reduce both the in-band and the out-of-band distortion per RF chain. This is the main scope of this paper.

Digital predistortion has been studied in the massive MIMO context in \cite{DPD_MM_1,DPD_MM_2,DPD_MM_3,DPD_MM_4}. In \cite{DPD_MM_1}, fully digital beamforming was assumed, and therefore a dedicated DPD unit for each transmitter is required. In \cite{DPD_MM_2,DPD_MM_3,DPD_MM_4}, DPD in hybrid MIMO was investigated. To this end, as the predistorter is operating in the digital baseband, a single DPD should linearize all the $M$ PAs simultaneously. This is essentially an underdetermined problem and will commonly lead to reduced linearization performance for the individual PAs. In \cite{DPD_MM_2}, the DPD learning was based on measuring only one of the PAs, while in \cite{DPD_MM_3}, the PAs per sub-array were assumed to be identical. However, these approaches will work satisfactorily only if the PA nonlinear characteristics are very similar - an assumption that is commonly far from practical. In \cite{DPD_MM_4}, a single DPD per sub-array was proposed based on the direct learning approach. The learning criteria is based on minimizing the sum of squared errors between the input and output signals of the PAs while using a dedicated observation RX chain per PA. The work in \cite{DPD_MM_4} was shown to provide better results compared to estimating the DPD parameters using only one of the PA elements. However, only memoryless DPD processing was proposed in \cite{DPD_MM_4} and therefore it was only tested using memoryless PAs, which is not a realistic case, especially when considering relatively wide-band transmit signals with tens or hundreds of MHz bandwidth.

In this paper, we propose a new structure for DPD learning in hybrid MIMO transmitters, which is both simpler and more effective than the current state-of-the-art. We argue that, because the individual PAs can anyway not be linearized perfectly, the objective should be to primarily reduce the distortions in the direction of the intended receiver. For the other spatial directions, \cite{OOB_Mollen} showed that the out-of-band emissions will be diluted due to non-coherent superposition of the transmit signals. This philosophy leads us to use the superposition signal of the individual PA outputs for DPD learning, and thus using only a single observation RX chain. In terms of main beam linearization, the proposed DPD is shown to give superior results compared to using only a single PA for learning. \textcolor{black}{To assess how the emissions behave in other spatial directions under the proposed DPD solution, we apply a similar numerical approach as \cite{OOB_Mollen}. Our results indicate that while the proposed DPD significantly reduces the unwanted emissions in the main beam direction, the out-of-band emissions in the other spatial directions are also well-behaving and essentially below those of the reference single-antenna transmitter due to the combined effects of DPD and beamforming.} The sensitivity of the technique to amplitude and phase imbalance between the feedback paths is also analyzed, and the effects are shown to be negligible with realistic imbalance values. Moreover, the proposed DPD structure and learning are developed taking into consideration the unavoidable memory effects in the PAs, \textcolor{black}{and can in general be adopted at below 6 GHz bands as well as at mmWave frequencies}. For realistic performance assessment, the proposed DPD is tested and evaluated using realistic PA models with memory which are extracted from actual hardware equipment that can be used in a real massive MIMO base station\footnote{Lund University Massive MIMO testbed, {\color{blue}\url{http://www.eit.lth.se/mamitheme}} {\color{black}The PA models are available through IEEEXplore.}}.

The rest of the article is structured as follows. Section \ref{sec:modeling} analyzes the nonlinear distortion created by the PAs of a single sub-array in the direction of the intended RX. Section \ref{sec:DPD_structure_learning} then introduces the proposed DPD structure and parameter learning method. The impacts of amplitude and phase mismatches between the feedback paths are then analyzed in Section \ref{sec:nonidealities}. Section \ref{sec:Simulations} presents some realistic simulation results using practical RF measurement based PA models with memory. Finally, Section \ref{sec:Conclusions} concludes the findings of this paper.

\section{Modeling and Measuring Nonlinear Distortion in Hybrid MIMO TX Sub-arrays}
\label{sec:modeling}
In this section, the nonlinear distortion due to the PAs in a hybrid MIMO transmitter architecture is analyzed as a first essential step towards developing an efficient DPD structure and parameter learning solution. 
Considering a single sub-array as shown in Fig. \ref{fig:TX_subarray_main_path}, yet without any DPD processing (i.e., $\tilde{x}(n) = x(n)$), the baseband equivalent input and output signals of the $m^{th}$ PA, assuming a parallel Hammerstein (PH) PA model \cite{PH_PA_Ref,ComplexityAnalysis,GhannouchiDec.2009}, read
\begin{align}
x_m(n) &= w_m x(n) \label{eq:PA_in_m} \\
y_m(n) &= \sum_{\substack{p=1 \\ p, \text{odd}}}^{P} f_{m,p,n} \star |x_m(n)|^{p-1} x_m(n), \label{eq:PA_out_m}
\end{align}
where $x(n)$ denotes the baseband equivalent transmit signal of the considered sub-array, $x_m(n)$ and $y_m(n)$ refer to the baseband equivalent input and output signals of the PA unit in the $m^{th}$ antenna branch while $w_m$ denotes the corresponding beamforming coefficient. 
Furthermore, $f_{m,p,n}$ denotes the $p^{th}$ order PH branch filter impulse response for the PA unit of the antenna branch $m$, and $\star$ is the convolution operator which is defined as $f_{m,p,n} \star x_m(n) = \sum_{\substack{l=0}}^{N} f_{m,p,l} x_m(n-l)$, where $N$ is the filter memory order. Assuming $|w_m| = 1$, i.e., only phase rotations are performed in the analog beamforming stage, the transmit signal of the $m^{th}$ antenna element can be equivalently re-written as
\begin{align}
y_m(n) &= w_m \sum_{\substack{p=1 \\ p, \text{ odd}}}^{P} f_{m,p,n} \star |x(n)|^{p-1} x(n). \label{eq:PA_out_m_new}
\end{align}

In general, the beamforming coefficients $w_m$ are chosen such that most of the allocated power is radiated towards the intended RX direction. 
Therefore, in order to further analyze the harmful radiated emissions, we primarily consider the nonlinear distortion which is radiated from the TX array towards the intended RX \cite{OOB_Mollen}. 
Assuming next, for simplicity, ideal beamforming in free-space line-of-sight (LoS) conditions, the individual signals $y_m(n)$ will coherently add up to construct an equivalent received or observed signal $y(n)$ of the form
\begin{align}
y(n) &= \sum_{\substack{m=1}}^{M} w_m^* y_m(n) \\
     &= \sum_{\substack{m=1}}^{M} \sum_{\substack{p=1 \\ p, \text{ odd}}}^{P} f_{m,p,n} \star |x(n)|^{p-1} x(n). \label{eq:PA_out}
\end{align}

In this work, we consider the out-of-band (OOB) emissions in the worst case scenario when the victim RX lies in the same direction as the intended RX, as discussed also in \cite{OOB_Mollen} and in the Introduction. In such scenarios, when assuming a single user per sub-array, the OOB emissions are similar to the classical emission scenarios and can be quantified using the adjacent channel leakage ratio (ACLR) metric. However, the exact method of evaluating the ACLR in large-array transmitters has not been decided yet in, e.g., 3GPP mobile radio network standardization. In this work we measure the ACLR based on the effective combined signal $y(n)$, which is essentially the sum of outputs from all the antenna elements per sub-array in the intended RX direction.
On the other hand, the in-band distortion of the effective radiated signal $y(n)$ will be very similar to the classical scenarios and will be quantified using the error vector magnitude (EVM) metric in this work. \textcolor{black}{Finally, we note that while the basic modeling and DPD processing developments in this article build on the PH or memory polynomial (MP) based approach, also more elaborate nonlinear models such as the generalized memory polynomial (GMP) \cite{GMP_morgan} can be adopted in a straight-forward manner.}

\begin{figure}[t!]
\centering
\includegraphics[width=1\linewidth]{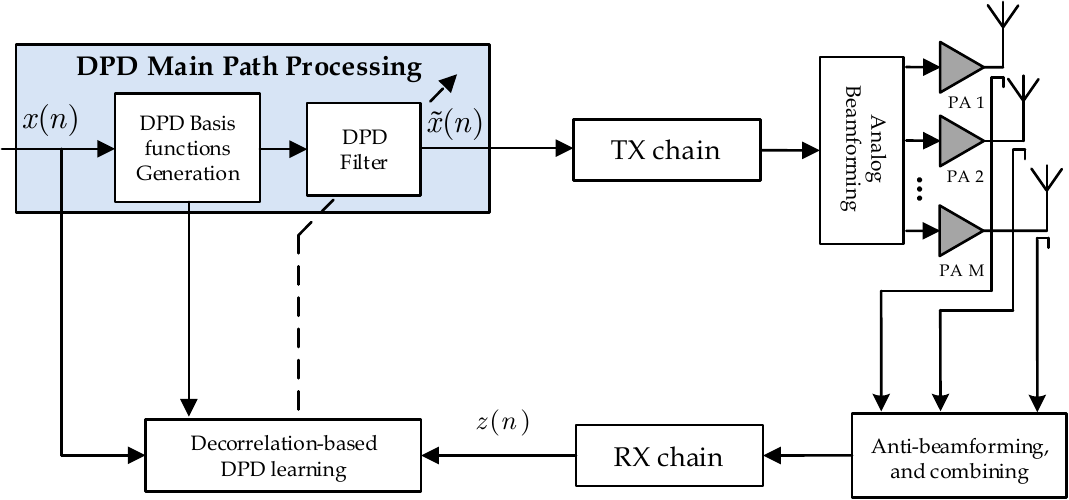}
\caption[]{Block diagram of the proposed DPD system for one sub-array.}
\label{fig:TX_subarray_DPD}
\end{figure}

\section{Proposed DPD Structure and Parameter Learning Solution}
\label{sec:DPD_structure_learning}
In the hybrid MIMO architecture, each sub-array is fed by a single RF upconversion chain. This implies that only a single DPD stage can be used per sub-array, as shown in Fig. \ref{fig:TX_subarray_DPD}. From one side, this reduces the complexity of the overall transmitter system in terms of the DPD processing, while on the other side it makes the linearization problem much more challenging, both from the DPD main path processing structure and the parameter learning perspectives, as the exact characteristics of the involved $M$ parallel PAs are generally different.

\subsection{Proposed DPD Structure}
Based on the nonlinear distortion analysis in the previous section, we formulate the proposed DPD structure and learning philosophy in this section keeping in mind that the main objective is to primarily minimize the harmful emissions in the intended RX direction, i.e., the in-band and OOB nonlinear distortion products in the effective combined signal $y(n)$ expressed in (\ref{eq:PA_out}).

We first rewrite (\ref{eq:PA_out}) such that the linear and nonlinear terms are separated as follows
\begin{align}
y(n) &= \sum_{\substack{m=1}}^{M} f_{m,1,n} \star x(n) + \sum_{\substack{m=1}}^{M} \sum_{\substack{p=3 \\ p, \text{ odd}}}^{P} f_{m,p,n} \star |x(n)|^{p-1} x(n) \\
&\color{black}{= f_{tot,1,n} \star x(n) + \sum_{\substack{p=3 \\ p, \text{ odd}}}^{P} f_{tot,p,n} \star |x(n)|^{p-1} x(n),}
\label{eq:PA_out_2}
\end{align}
\textcolor{black}{where $f_{tot,p,n} = \sum_{\substack{m=1}}^{M} f_{m,p,n}$.} From (\ref{eq:PA_out_2}), it can be seen that the nonlinear term of $y(n)$ is composed of a linear combination of the static nonlinear (SNL) basis functions $u_p(n) = |x(n)|^{p-1} x(n)$ and their delayed replicas. \textcolor{black}{Furthermore, the effective branch filters $f_{tot,p,n}$ depend only on the sums of the individual PA model branch filters $f_{m,p,n}, m = 1, 2,\hdots, M$.} 
In general, we focus our attention mostly on the nonlinear distortion, since the linear distortion term in (\ref{eq:PA_out_2}) is anyway usually equalized at the receiver side and can thus be considered to be part of the overall wireless communications channel.
Consequently, the key idea of the proposed DPD structure is to inject a proper additional low-power cancellation signal, with structural similarity to the nonlinear terms in (\ref{eq:PA_out_2}), at the input of the PAs of the considered sub-array such that the radiated in-band and OOB nonlinear distortion products are minimized in the intended RX direction.

Stemming from the above modeling, an appropriate digital injection signal can be obtained by adopting the SNL basis functions $u_q(n), q = 3, 5,\hdots, Q,$ combined with proper filtering using a bank of DPD filters, $\alpha_{q,n}$, with memory order $N_q$. 
In general, incorporating such DPD processing with polynomial order $Q$, the output signal of the DPD processing stage of the considered sub-array reads
\begin{align}
\tilde{x}(n) =  x(n) + \sum_{\substack{q=3\\ q, \text{odd}}}^{Q} \alpha^*_{q,n} \star  u_{q}(n).
\label{eq:PA_In_with_DPD}
\end{align}
Here, and in the continuation, we use $\tilde{(.)}$ variables to indicate DPD-based processing and corresponding signals.
The achievable suppression of the nonlinear distortion depends directly on the selection and optimization of the DPD filter coefficients $\alpha_{q,n}$. This is addressed in detail in the next subsection. We also note that an additional branch filter can be applied to the linear signal term in (\ref{eq:PA_In_with_DPD}) if, e.g., linear response pre-equalization is pursued.

\subsection{Proposed Combined Feedback based DPD Learning}
\label{sec:DPD_learning}
The main philosophy of the proposed DPD learning is to minimize the correlation between the nonlinear distortion radiated from the considered sub-array and the SNL basis functions $u_q(n)$ and their delayed replicas. Once such correlation minimization is achieved, the level of the nonlinear distortion is significantly reduced. This type of a decorrelation-based learning criteria has been introduced earlier by the authors in \cite{DPD_MM_1,Concurrent_abdelaziz} in the context of single-antenna transmitters. In this article, a similar approach is adopted and developed in the context of DPD parameter learning in hybrid MIMO transmitters.


In order to extract the effective nonlinear distortion at the sub-array output, while also using only a single RX chain in the observation system, we propose to explicitly combine the individual outputs of each PA per sub-array. This can be realized by using $M$ directional couplers followed by a co-phasing (or ``anti-beamforming") and combining module before the feedback RX chain as shown in Fig. \ref{fig:TX_subarray_DPD}. The purpose of the anti-beamforming is to counteract the effect of the beamforming coefficients in the analog beamforming module such that the observed signal in the feedback observation RX corresponds to the actual signal radiated in the intended RX direction ($y(n)$). 
Another practical alternative is to momentarily set all beamforming weights to one ($w_m=1$ for all $m$), for the period of DPD parameter learning, and then simply sum up the individual PA output signals, which essentially yields the same observation waveform.

Consequently the baseband equivalent observation signal at the feedback RX output, denoted by $z(n)$, while assuming that $|w_m| = 1$, reads  
\begin{align}
z(n) &= \sum_{\substack{m=1}}^{M} w^*_m \textcolor{black}{g_c} y_m(n) \\
     &= \textcolor{black}{g_c}\sum_{\substack{m=1}}^{M} |w_m|^2 \sum_{\substack{p=1 \\ p, \text{ odd}}}^{P} f_{m,p,n} \star |x(n)|^{p-1} x(n) \\ 
		 &= \textcolor{black}{g_c}\sum_{\substack{m=1}}^{M} \sum_{\substack{p=1 \\ p, \text{ odd}}}^{P} f_{m,p,n} \star |x(n)|^{p-1} x(n),
		\label{eq:PA_observation}
\end{align}
\textcolor{black}{where $g_c$ denotes the coupling factor in the individual feedback paths. For presentation simplicity, $g_c$ is assumed above to be identical in all feedback branches. Practical mismatches between the feedback branches will then be considered in detail in Section \ref{sec:nonidealities} as well as in the numerical experiments in Section \ref{sec:Simulations}. Notice also that since the anti-beamforming stage cancels or removes the effects of the specific beamforming coefficients, parameter learning can take place during the normal operation of the transmitter. Alternatively, a dedicated learning period can also be adopted.} 

\textcolor{black}{In order to utilize the observation signal $z(n)$ in the DPD learning, we can rewrite (\ref{eq:PA_observation}) as 
\begin{align}
z(n) = G x(n) + d(n), 
\end{align}
where $G$ is the effective complex linear gain while $d(n)$ corresponds to the total effective distortion signal due to the PA units.} The actual error signal which is then used for the decorrelation-based parameter learning is calculated as follows
\begin{align}
e(n) = z(n) - \textcolor{black}{\hat{G}} x(n), 
\label{eq:total_error}
\end{align}
\textcolor{black}{where $\hat{G}$ is the effective linear gain estimate which can be obtained in practice by using, e.g., block least squares (LS).}
This error signal seeks to provide information at waveform level about the currently prevailing nonlinear distortion samples in the effective combined signal relative to the ideal signal samples $x(n)$. In cases where there is substantial frequency-selectivity in the effective linear response, an actual multitap filter can be estimated and utilized in (\ref{eq:total_error}).

In general, the SNL basis functions $u_q(n) = |x(n)|^{q-1} x(n)$ and their delayed replicas are strongly mutually correlated, and thus basis function orthogonalization is required in order to have a faster and smoother convergence of the DPD parameter estimates during the learning process \cite{Haykin_AdaptiveFilters}. In principle, any suitable orthogonalization/whitening transformation with a triangular orthogonalization matrix can be adopted, e.g., QR decomposition (Gram-Schmidt type) or one based on Cholesky decomposition of the covariance matrix of the basis functions. For clarity, the orthogonalized basis functions are denoted in the following by $s_q(n)$.

The actual block-adaptive decorrelation-based DPD coefficient update, with learning rate $\mu$, then reads
\begin{align}
\bar{\boldsymbol{\alpha}}(i+1) = \bar{\boldsymbol{\alpha}}(i) -  \mu \: [\textbf{e}(i)^H \textbf{S}(i)]^T, 
\label{eq:BlockAdaptive}
\end{align}
where $\textbf{e}(i) = [e(n_i) \:\: ... \:\: e(n_i+B-1)]^T$ is a block of $B$ observed samples of the error signal $e(n)$, while $n_i$ denotes the index of the first sample within block $i$. 
Furthermore, $\bar{\boldsymbol{\alpha}}(i)$ in (\ref{eq:BlockAdaptive}) is defined as follows
\begin{align}
\boldsymbol{\alpha}_q(i) &= [\alpha_{q,0}(i)\:\alpha_{q,1}(i)\: ... \: \alpha_{q,N_q}(i)]^T \\
\bar{\boldsymbol{\alpha}}(i) &= [\boldsymbol{\alpha}_{3}(i)^T \: \boldsymbol{\alpha}_{5}(i)^T \: ... \: \boldsymbol{\alpha}_{Q}(i)^T]^T,
\end{align}
while $\textbf{S}(i)$ in (\ref{eq:BlockAdaptive}) is defined using the orthogonal basis function samples $s_q(n)$ as
\begin{align}
\textbf{s}_{q}(n_i) &= [s_{q}(n_i)\: ... \: s_{q}(n_i-N_q)] \\
\textbf{S}_{q}(i) &= [\textbf{s}_{q}(n_i)^T \: ... \: \textbf{s}_{q}(n_i+B-1)^T]^T \\
\textbf{S}(i) &= [\textbf{S}_{3}(i) \:\textbf{S}_{5}(i)\: ... \: \textbf{S}_{Q}(i)].  
\end{align}
The updated DPD coefficients $\bar{\boldsymbol{\alpha}}(i+1)$ are then used to filter the next block of $B$ samples, and the process is iterated until convergence. 
Using the adaptive filter update in (\ref{eq:BlockAdaptive}), it can be shown that the learning algorithm converges to the point where the residual nonlinear distortion becomes uncorrelated with the adopted orthogonalized basis functions, and hence the name decorrelation-based learning.

\textcolor{black}{Finally, we note that the above decorrelation-based iterative learning rule can be either executed during the specific parameter learning/calibration periods, or alternatively, be even running continuously, in conjunction with the anti-beamforming based combined feedback signal, if one wishes to track the potential changes in the PA characteristics as accurately as possible. This is because even though the served intended receiver(s) change in practical radio networks commonly at 1 ms time scale (the scheduling interval in LTE/LTE-Advanced mobile radio networks), or even faster, the anti-beamforming stage makes the feedback signal independent of the actual value of the intended receiver direction, and hence the algorithm can run continuously.}

\section{Impact of Mismatches Between the Feedback Branches}
\label{sec:nonidealities}
In order to obtain the feedback signal $z(n)$ which is used for the DPD learning, the outputs of the individual PAs are first extracted using $M$ directional couplers, then co-phased and combined in the analog domain before being applied into a single observation RX chain which brings the observation signal down to baseband, as shown in Fig. \ref{fig:TX_subarray_DPD}. Consequently, in the actual physical circuit implementation, there can be amplitude and phase mismatches between the $M$ observation branches prior to and within combining. In the following, these mismatches are analyzed and their effect on the proposed DPD system and its performance is discussed.

Assuming $\textcolor{black}{\epsilon_m} = \beta_m e^{j\phi_m}$ denotes the complex gain deviation relative to the nominal coupling factor $\textcolor{black}{g_c}$ in the $m^{th}$ feedback coupler path, where $\beta_m$ and $\phi_m$ are the corresponding gain and phase mismatches, the baseband equivalent combined observation signal $z(n)$ then reads
\begin{align}
z(n) &= \sum_{\substack{m=1}}^{M} \textcolor{black}{g_c(1 + \epsilon_m)} \sum_{\substack{p=1 \\ p, \text{ odd}}}^{P} f_{m,p,n} \star |x(n)|^{p-1} x(n) \\
&= \textcolor{black}{g_c}\sum_{\substack{m=1}}^{M} \sum_{\substack{p=1 \\ p, \text{ odd}}}^{P} f_{m,p,n} \star |x(n)|^{p-1} x(n) \nonumber\\
&+ \textcolor{black}{g_c}\sum_{\substack{m=1}}^{M} \textcolor{black}{\epsilon_m} \sum_{\substack{p=1 \\ p, \text{ odd}}}^{P} f_{m,p,n} \star |x(n)|^{p-1} x(n).
		\label{eq:PA_observation_error}
\end{align}
When $\textcolor{black}{|\epsilon_m|} \ll 1$, the second term in (\ref{eq:PA_observation_error}) can be essentially neglected and we return back to the expression in (\ref{eq:PA_observation}). However, when the gain and phase mismatches start to increase, the combined observation signal starts to gradually degrade. Meanwhile, the assumption that the combined observation signal $z(n)$ is composed of a linear combination of the SNL basis functions and their delayed replicas will still hold. 

In order to more explicitly analyze the impact of such gain and phase mismatches between the feedback branches on the DPD learning, and consequently the DPD performance, in closed-form, we proceed as follows. For mathematical tractability, we assume simple third-order memoryless processing in both the PA and DPD models. 
For reference, we first derive the optimum decorrelation-based DPD coefficient without any mismatches, being then followed by the corresponding optimum coefficient derivation under the branch mismatches. This allows us to analytically address how much the mismatches affect or bias the DPD coefficient, in the simple example case of a third-order DPD system.

Now, the DPD output signal $\tilde{x}(n)$ with third-order memoryless DPD processing reads
\begin{align}
\tilde{x}(n) =  x(n) + \alpha^*_{3} |x(n)|^2 x(n),
\label{eq:analysis1}
\end{align}
where $\alpha_{3}$ is the third-order DPD coefficient applied in its conjugated form in order to conform with the notation adopted in Section \ref{sec:DPD_structure_learning}.  
The corresponding output of the $m^{th}$ PA unit with DPD becomes
\begin{align}
\tilde{y}_m(n) &= w_m [f_{m,1} \tilde{x}(n) + f_{m,3} |\tilde{x}(n)|^2 \tilde{x}(n)]. 
\end{align}
Then, the combined observation at the feedback receiver output, with DPD included, reads
\begin{align}
\tilde{z}(n) &= \sum_{\substack{m=1}}^{M} w^*_m \textcolor{black}{g_c} \tilde{y}_m(n) \\
             &= \textcolor{black}{g_c}\sum_{\substack{m=1}}^{M} (f_{m,1} \tilde{x}(n) + f_{m,3} |\tilde{x}(n)|^2 \tilde{x}(n)).
\label{eq:analysis2}
\end{align}
Substituting (\ref{eq:analysis1}) into (\ref{eq:analysis2}) yields
\begin{align}
\tilde{z}(n) &= x(n) \textcolor{black}{g_c}\sum_{\substack{m=1}}^{M} f_{m,1} \nonumber\\
&+ |x(n)|^2 x(n) \textcolor{black}{g_c}\sum_{\substack{m=1}}^{M} (\alpha^*_{3} f_{m,1} + f_{m,3})  \nonumber\\ 
&+ |x(n)|^4 x(n) \textcolor{black}{g_c}\sum_{\substack{m=1}}^{M} (\alpha_{3} +  2\alpha^*_{3})f_{m,3}  \nonumber\\ 
&+ |x(n)|^6 x(n) \textcolor{black}{g_c}\sum_{\substack{m=1}}^{M} (2|\alpha_{3}|^2 +  \alpha^{*2}_{3})f_{m,3} \nonumber\\ 
&+ |x(n)|^8 x(n) \textcolor{black}{g_c}\sum_{\substack{m=1}}^{M} \alpha^*_{3}|\alpha_{3}|^2 f_{m,3}. 
\label{eq:analysis3}
\end{align}
Since the decorrelation-based learning algorithm aims at minimizing the correlation between the error signal observed at feedback receiver output, i.e., $e(n) = \tilde{z}(n) - \hat{G} x(n)$, and the SNL third-order basis function $|x(n)|^2 x(n)$, we evaluate the expression of this correlation $\E \left[|x(n)|^2 x^*(n) e(n)\right]$ in closed-form \textcolor{black}{while assuming a perfect estimate of the effective linear gain $G$.}
We first write
\begin{align}
&\E \left[|x(n)|^2 x^*(n) e(n)\right] =  
 \E \left[|x(n)|^6 \textcolor{black}{g_c}\sum_{\substack{m=1}}^{M} (\alpha^*_{3} f_{m,1} + f_{m,3})\right] \nonumber\\ 
&+ \E \left[|x(n)|^8 \textcolor{black}{g_c}\sum_{\substack{m=1}}^{M} (\alpha_{3} +  2\alpha^*_{3})f_{m,3}\right] \nonumber\\
&+ \E \left[|x(n)|^{10} \textcolor{black}{g_c}\sum_{\substack{m=1}}^{M} (2|\alpha_{3}|^2 +  \alpha^{*2}_{3})f_{m,3}\right] \nonumber\\
&+ \E \left[|x(n)|^{12} \textcolor{black}{g_c}\sum_{\substack{m=1}}^{M} \alpha^*_{3}|\alpha_{3}|^2 f_{m,3}\right].
\label{eq:analysis4}
\end{align}
In order to analytically calculate the DPD coefficient $\alpha_3$ that minimizes $\E \left[|x(n)|^2 x^*(n) e(n)\right]$, we set (\ref{eq:analysis4}) to zero. While neglecting the higher-order terms by assuming that they are vanishingly small, \textcolor{black}{as $\alpha_3$ is in general a small number with any reasonable PA nonlinear response characteristics,} the correlation minimization approach yields
\textcolor{black}{
\begin{align}
&\alpha^*_{3,opt} \E |x(n)|^6 \textcolor{black}{g_c} \sum_{\substack{m=1}}^{M} f_{m,1} \nonumber\\ 
&+ (\alpha_{3,opt} +  2\alpha^*_{3,opt}) \E |x(n)|^8 \textcolor{black}{g_c}\sum_{\substack{m=1}}^{M} f_{m,3} \nonumber\\
&= - \E|x(n)|^6 \textcolor{black}{g_c}\sum_{\substack{m=1}}^{M} f_{m,3}.
\label{eq:analysis5}
\end{align}
}
\textcolor{black}{Then, denoting $\sum_{\substack{m=1}}^{M} f_{m,3} / \sum_{\substack{m=1}}^{M} f_{m,1}$ by $F_{31}$ we get the following expression}
\textcolor{black}{
\begin{align}
\alpha^*_{3,opt} = -F_{31} \left[1 + (\alpha_{3,opt} +  2\alpha^*_{3,opt}) \frac{\E |x(n)|^8}{\E |x(n)|^6} \right].
\label{eq:analysis6}
\end{align}
}
Taking the complex conjugate of (\ref{eq:analysis6}) provides us with a second equation in $\textcolor{black}{\alpha_{3,opt}}$ and $\textcolor{black}{\alpha^*_{3,opt}}$ which reads
\textcolor{black}{
\begin{align}
\alpha_{3,opt} = -F_{31}^* \left[1 + (\alpha^*_{3,opt} +  2\alpha_{3,opt}) \frac{\E |x(n)|^8}{\E |x(n)|^6} \right].
\label{eq:analysis62}
\end{align}
}
The expressions in (\ref{eq:analysis6}) and (\ref{eq:analysis62}) allow then solving for $\textcolor{black}{\alpha_{3,opt}}$ which yields
\textcolor{black}{
\begin{align}
\alpha_{3,opt} = \frac{-F_{31}^*(1 + F_{31} \E_{86})}{3|F_{31}|^2 \E_{86}^2 + 2 \E_{86} (F_{31} + F_{31}^*) + 1},
\label{eq:analysis7}
\end{align}
}
where $\E_{86} = \frac{\E |x(n)|^8}{\E |x(n)|^6}$. This expression serves as reference and comparison point for addressing the branch mismatch impact.

\textcolor{black}{Next, we introduce amplitude and phase mismatches in the feedback coupling paths and re-derive the expression for $\alpha_{3,opt}$ in order to examine the effect of such mismatches on the proposed learning algorithm. The optimum DPD coefficient with mismatches included is denoted by $\bar{\alpha}_{3,opt}$, for notational clarity.}
The feedback observation signal $\tilde{z}(n)$, with mismatches included, now reads
\begin{align}
\tilde{z}(n) = \textcolor{black}{g_c}\sum_{\substack{m=1}}^{M} \textcolor{black}{(1+\epsilon_m)}(f_{m,1} \tilde{x}(n) + f_{m,3} |\tilde{x}(n)|^2 \tilde{x}(n)).
\end{align}
Performing similar analysis steps as above, we get the following expression for the decorrelation-based optimum DPD coefficient $\bar{\alpha}_{3,opt}$, expressed as
\begin{align}
\bar{\alpha}^*_{3,opt} &= -\frac{\sum_{\substack{m=1}}^{M} f_{m,3}(1+\epsilon_m)}{\sum_{\substack{m=1}}^{M} f_{m,1}(1+\epsilon_m)} \nonumber\\ 
&\times \left[1 + (\bar{\alpha}_{3,opt} +  2\bar{\alpha}^*_{3,opt}) \frac{\E |x(n)|^8}{\E |x(n)|^6} \right].
\label{eq:analysis11}
\end{align}
\textcolor{black}{Then, denoting $\sum_{\substack{m=1}}^{M} f_{m,3}(1+\epsilon_m) / \sum_{\substack{m=1}}^{M} f_{m,1}(1+\epsilon_m)$ by $\bar{F}_{31}$ we get the following expression}
\textcolor{black}{
\begin{align}
\bar{\alpha}^*_{3,opt} = -\bar{F}_{31} \left[1 + (\bar{\alpha}_{3,opt} +  2\bar{\alpha}^*_{3,opt}) \frac{\E |x(n)|^8}{\E |x(n)|^6} \right].
\label{eq:analysis12}
\end{align}
}
Then, using similar analysis steps as in the case without mismatches, \textcolor{black}{the expression for $\bar{\alpha}_{3,opt}$ with mismatches can be shown to read}
\textcolor{black}{
\begin{align}
\bar{\alpha}_{3,opt} = \frac{-\bar{F}_{31}^*(1 + \bar{F}_{31} \E_{86})}{3|\bar{F}_{31}|^2 \E_{86}^2 + 2 \E_{86} (\bar{F}_{31} + \bar{F}_{31}^*) + 1},
\label{eq:analysis13}
\end{align}
}
\textcolor{black}{where $\bar{F}_{31}$ is given by}
\textcolor{black}{
\begin{align}
\bar{F}_{31} = \frac{\sum_{\substack{m=1}}^{M} f_{m,3}+\sum_{\substack{m=1}}^{M} f_{m,3}\epsilon_m}{\sum_{\substack{m=1}}^{M} f_{m,1}+\sum_{\substack{m=1}}^{M} f_{m,1}\epsilon_m}.
\end{align}
}
\textcolor{black}{
When using a relatively large number of antennas per sub-array, $M$, then $\sum_{\substack{m=1}}^{M} f_{m,3}\epsilon_m\rightarrow c\E[f_{m,3}\epsilon_m]$
where $c$ is a scaling constant, and similarly $\sum_{\substack{m=1}}^{M} f_{m,1}\epsilon_m\rightarrow c\E[f_{m,1}\epsilon_m]$. Assuming $\epsilon_m$ is a random variable with zero mean, and since $\epsilon_m$ is, in general, independent of both $f_{m,3}$ and $f_{m,1}$, then $\bar{F}_{31} \approx F_{31}$, and consequently (\ref{eq:analysis13}) reduces to (\ref{eq:analysis7}). 
}
Thus, the analysis shows that the mismatches in the feedback branches have a very small effect on the proposed decorrelation-based DPD parameter learning, and consequently its performance. Thus, the proposed sub-array DPD system is robust against the possible feedback coupling branch mismatches, a finding that we also (re)confirm using the numerical experiments in the following section. \textcolor{black}{We note that while the analytical mismatch analysis above builds on the simplifying assumption of third-order memoryless models, our numerical experiments will include higher-order nonlinearities and memory in both the PA units as well as in the DPD processing stage.}


\section{Numerical Experiments}
\label{sec:Simulations}
\begin{figure}[t!]
\centering
\includegraphics[width=1\linewidth]{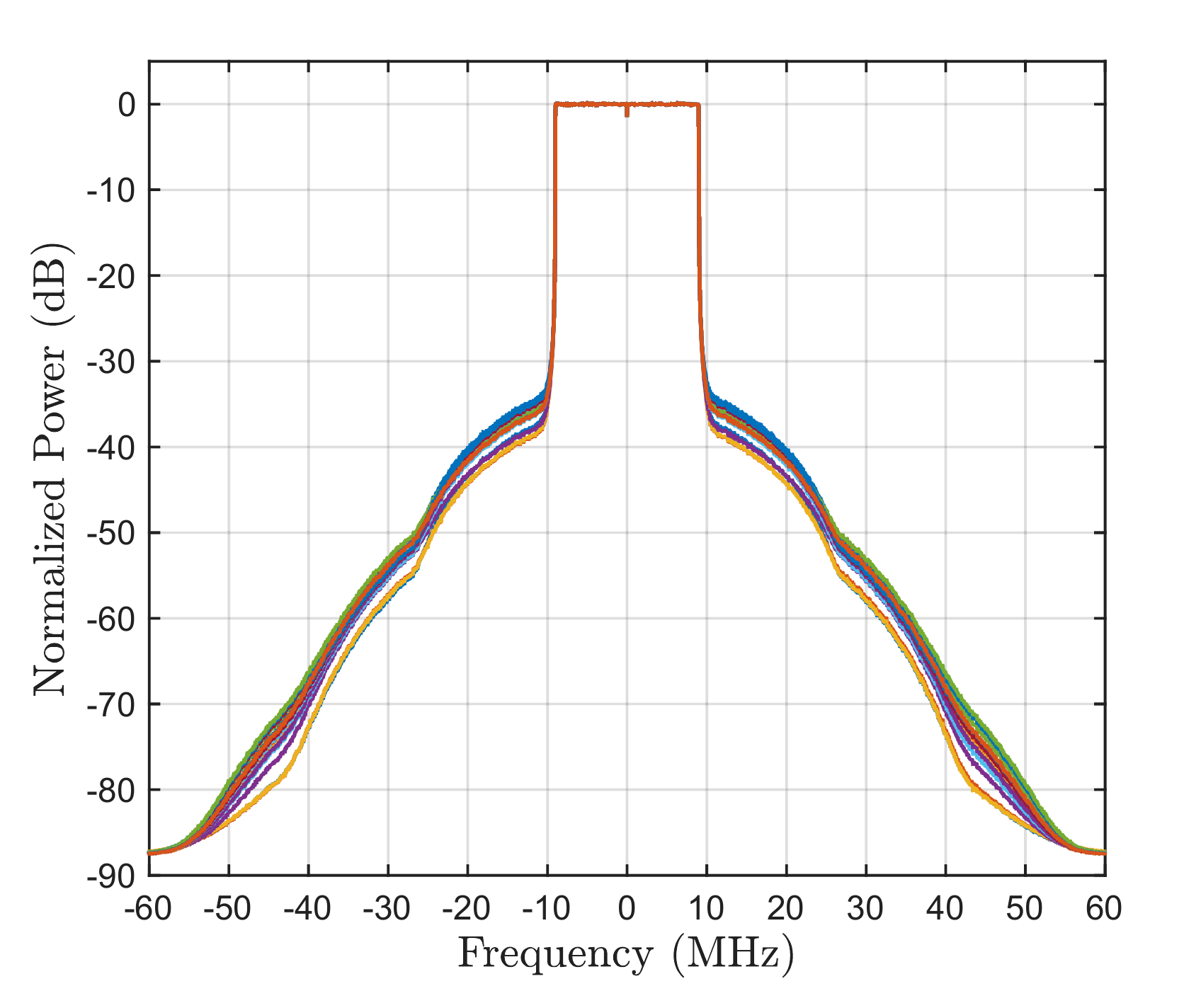}
\caption[]{Normalized individual PA output spectra of 16 different PA models extracted from a true large-array transmitter system at 120 MHz sample rate. The transmitted OFDM carrier is 20 MHz wide with 16-QAM subcarrier modulation, and the PAPR is 8.3 dB. An 11th-order PH model with memory is extracted per PA. The passband power of every PA model is normalized to 0 dB.}
\label{fig:PSD_16_PAs}
\end{figure}

In this section, a quantitative performance analysis of the proposed DPD solution is presented using comprehensive Matlab simulations with practical measured models for PAs with memory. 
The measured PA models are obtained from the Lund University massive MIMO test-bed which is one of the most established large-array transceiver platforms currently available, and includes 100 PA units overall.
The proposed DPD which uses the combined feedback signal is compared against a classical DPD approach which uses only a single PA output for learning. The PA models are 11th-order PH models extracted from individual USRP modules that are used in the Lund massive MIMO hardware testbed transmitting at 2 GHz RF frequency. The sample rate used to extract these models is 120 MHz. 
The credibility and practicability of the results presented in this section is thus high when compared to state-of-the-art works in DPD for hybrid MIMO transmitters which usually assume substantially more simple PA models without memory \cite{DPD_MM_4}, or even that all PA units in such array structure would be identical \cite{DPD_MM_2,DPD_MM_3}. 
The signal used in the PA measurements as well as in our DPD simulations is a 20 MHz OFDM signal with 16-QAM subcarrier modulation. Iterative clipping and filtering-based PAPR reduction is applied to the transmit signal limiting the actual PAPR of the signal to approximately 8.3 dB \cite{PAPR_Clipping_Filtering}. The output power spectra of 16 different PAs of representative nature are shown in Fig. \ref{fig:PSD_16_PAs}.
\textcolor{black}{\subsection{DPD Performance Results and Analysis}}
\label{subsec:DPD_Results}
First, we address the achievable linearization performance in the intended RX direction. Both the inband waveform purity and the adjacent channel interference due to spectral regrowth are quantified using the error vector magnitude (EVM), and the ACLR metrics, respectively \cite{3GPP_BS}. The EVM and ACLR are calculated for the effective combined signal in the intended RX direction as explained in the previous sections.
The EVM is defined as
\begin{equation}
EVM_{\%} = \sqrt{P_{error}/P_{ref}} \times 100\%,
\end{equation}
where $P_{error}$ is the power of the error signal, defined as the difference between the ideal symbol values and the corresponding symbol rate complex samples at the array output in the intended RX direction, both normalized to the same average power, while $P_{ref}$ is the reference power of the ideal symbol constellation. 
Typically in EVM evaluations, linear distortion of the transmit chain is equalized prior to calculating the error signal \cite{Dahlman4Gbook}, and this is also what we do in this work. 
In turn, the ACLR is defined as the ratio of the emitted powers within the wanted channel ($P_{wanted}$) and the adjacent channel ($P_{adjacent}$), respectively \cite{3GPP}, interpreted also for the effective combined signal in the direction of the intended RX, namely
\begin{equation}
ACLR_{dB} = 10 \log_{10} \frac{P_{wanted}}{P_{adjacent}}.
\end{equation}
In this work, the channel bandwidth of the wanted signal is defined as the bandwidth which contains 99\% of the total emitted power in the main beam direction. The adjacent channel measurement bandwidth is equal to this.

\begin{figure}[t!]
\centering
\includegraphics[width=1\linewidth]{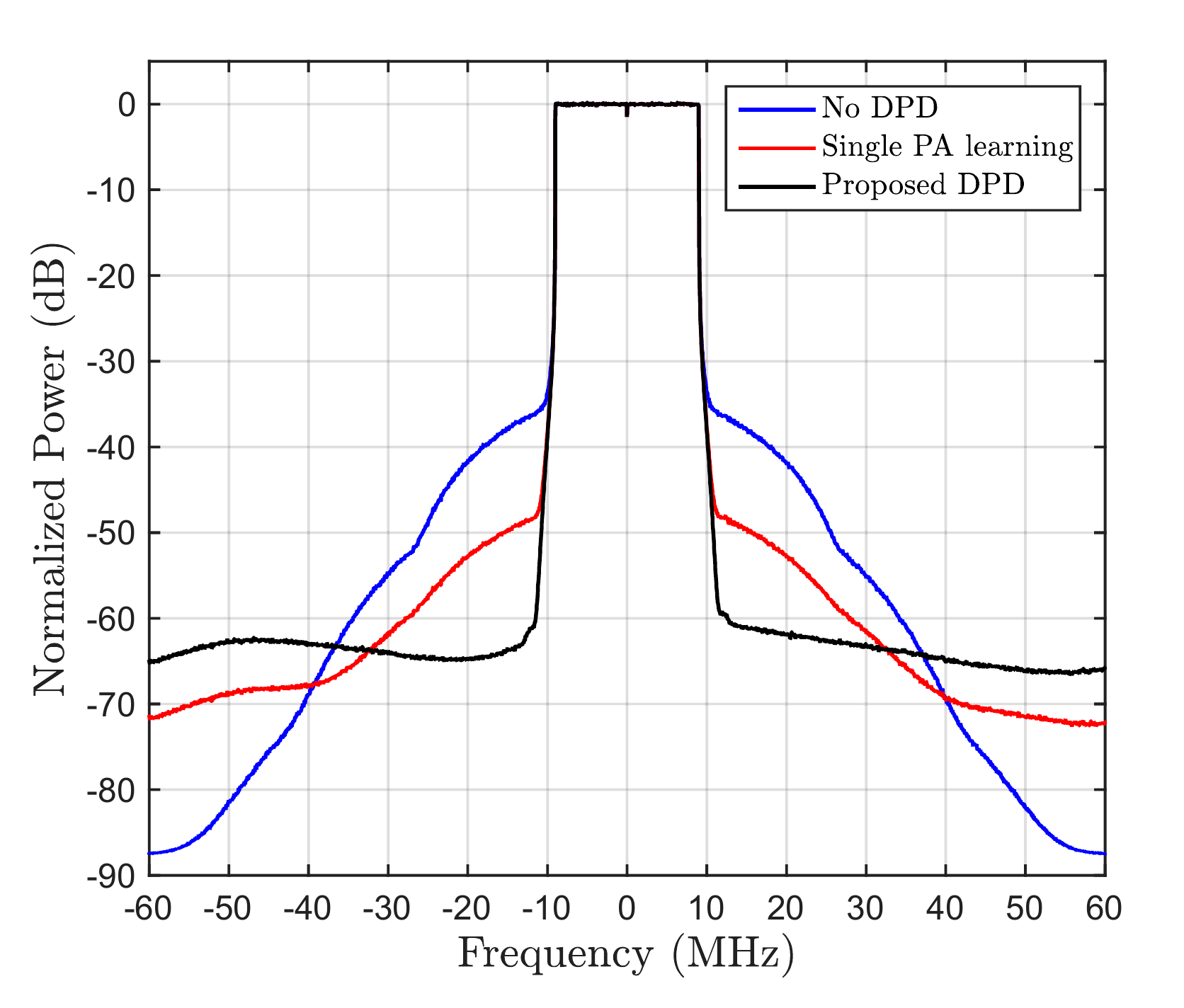}
\caption[]{Normalized output spectra of the effective combined signals from 16 PA elements in the direction of the intended receiver. Three scenarios are shown: without DPD, with DPD estimated for a single PA unit and applied to all PAs, and with the proposed DPD. The PA models are 11th-order PH models with memory extracted from a true large-array transmitter system at 120 MHz sample rate. The transmitted OFDM carrier is 20 MHz wide with 16-QAM subcarrier modulation and 8.3 dB PAPR. Amplitude mismatches between $-10$ and $10\%$ and phase mismatches between $-10$ and $10^o$ are incorporated in the feedback paths when using the proposed DPD.}
\label{fig:PSD_final}
\end{figure}

\begin{table}[t!]
\caption{EVM and ACLR results}
\centering
{\begin{tabular}{lll}\hline
                                  &\vline\:\:\: EVM ($\%$)  &\vline\:\:\: ACLR L\:/\:R (dBc)    \\\hline
Without DPD                       &\vline\:\:\: 3.17        &\vline\:\:\: 40.48\:/\:40.58       \\\hline
With single PA learning 				  &\vline\:\:\: 2.09        &\vline\:\:\: 52.01\:/\:51.91       \\\hline
With proposed DPD 							  &\vline\:\:\: 1.85        &\vline\:\:\: 63.63\:/\:61.42       \\\hline
\end{tabular}}{}
\label{tab:EVM_ACLR}
\end{table}

\begin{figure}[t!]
\centering
\centerline{\includegraphics[width=0.48\textwidth]{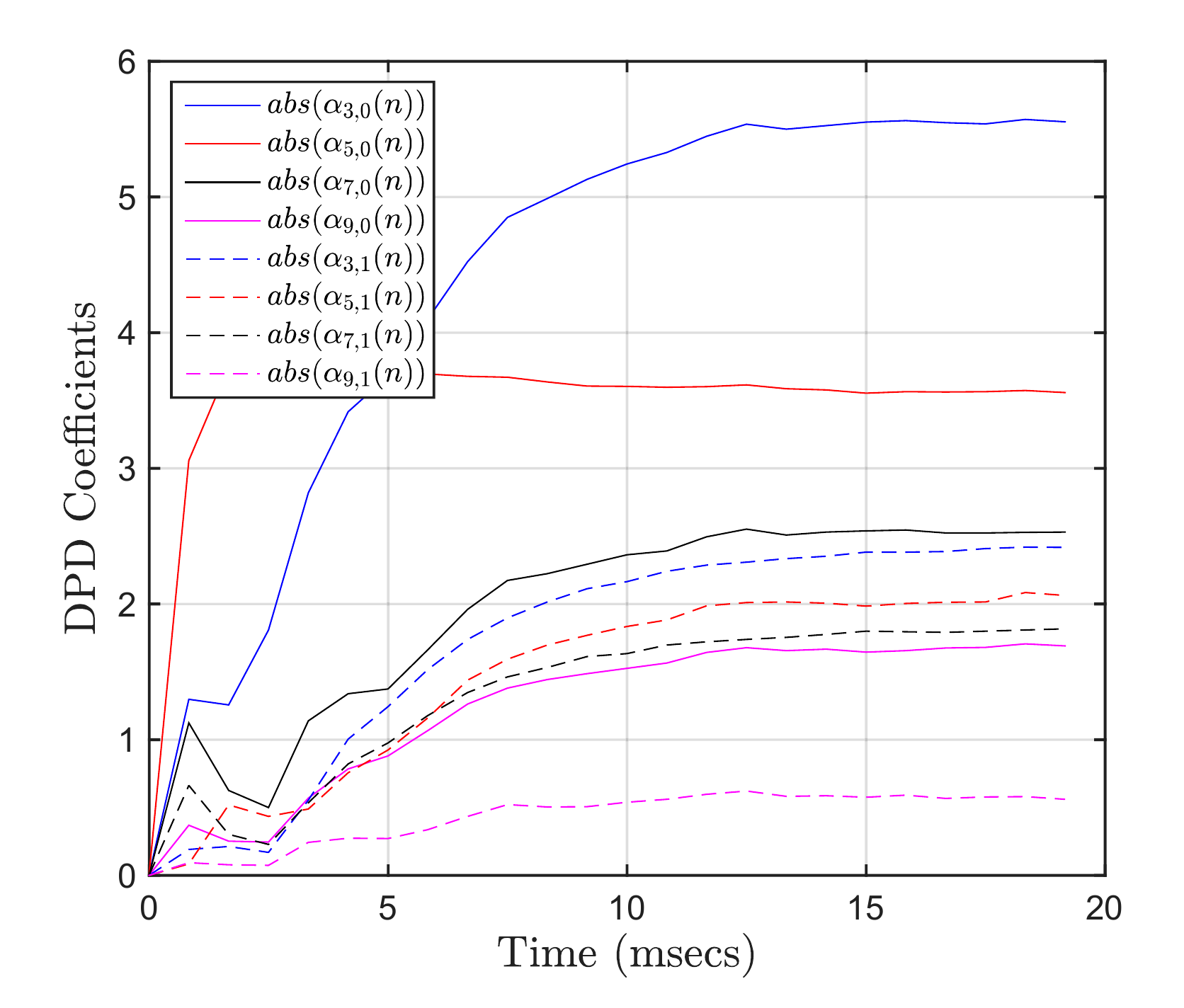}}
\caption[]{Example convergence of the first two memory taps, per basis function, of the proposed ninth-order decorrelation-based DPD using a single realization of a 20 MHz OFDM carrier with 16-QAM subcarrier modulation and 8.3 dB PAPR. Amplitude mismatches between $-10$ and $10\%$ and phase mismatches between $-10$ and $10^o$ are incorporated in the feedback paths when using the proposed DPD. The PA models are 11th-order PH models with memory, extracted from a true large-array transmitter system.}
\label{fig:DPD_coeffs}
\end{figure}

The nonlinearity order $Q$ of the proposed DPD is 9, and the DPD memory depth $N$ is equal to 3 (i.e., 4 memory taps per PH branch filter). The learning block-size $B$ used by the DPD is $100k$ samples, and 24 block adaptive iterations are used. These parameters are used both in the proposed DPD and in the reference DPD method which uses only a single PA for learning, while the considered sub-array size $M=16$. \textcolor{black}{The effective linear gain, $G$, is estimated using ordinary block least squares (LS), per each block iteration.} 

The power spectrum of the effective combined signal from 16 PA elements in the direction of the intended receiver is shown in Fig. \ref{fig:PSD_final} in three scenarios: without DPD, with decorrelation-based DPD estimated using the first PA only, and with the proposed DPD. 
Notice that we also implemented for reference the state-of-the-art method from \cite{DPD_MM_4} but since the method described in \cite{DPD_MM_4} does not take into account the PA memory, the resulting performance is not comparable at all to the other considered methods, and hence not included in the results.
Table \ref{tab:EVM_ACLR} shows the corresponding EVM and ACLR values showing an excellent linearization performance of the proposed DPD system. More than $10$ dB gain in ACLR is achieved when using the proposed DPD compared to using a single PA output for learning.
When using the proposed DPD, random amplitude and phase mismatches are included in the feedback paths to facilitate a realistic performance evaluation scenario. The amplitude mismatches are uniformly distributed between $-10$ and $10\%$, while the phase mismatches are uniformly distributed between $-10$ and $10^o$. Despite such relatively large feedback network mismatches, excellent linearization performance is obtained which verifies the analytical findings regarding the robustness against mismatches reported in Section \ref{sec:nonidealities}.
Fig. \ref{fig:DPD_coeffs} presents an example of the proposed DPD coefficient behavior, during the learning phase, while showing only the first two memory taps (out of four) per SNL basis function, to keep the visual illustration readable. It is clear from Fig. \ref{fig:DPD_coeffs} that the coefficients converge in a reliable and relatively fast manner, when compared to any practical or realistic potential rate of change of the characteristics of the PAs in the considered sub-array. Such good convergence properties are partly due to the basis function orthogonalization processing, as explained in section \ref{sec:DPD_learning}.

\textcolor{black}{\subsection{Analysis of Unwanted Emissions in Spatial Domain}
Next, we analyze how the inband power and out-of-band emissions, in all different spatial directions, behave after applying the proposed DPD. In \cite{OOB_Mollen}, it was shown that the OOB emissions of massive MIMO transmitters essentially follow the beam pattern of the array. Thus, OOB emissions are more powerful in the direction of the intended receiver, while other directions are attenuated. However, there are no studies that analyze how the OOB emissions of the array transmitter behave after applying a certain DPD solution. This analysis is of great importance, especially in the problem at hand, where the developed DPD algorithm primarily considers the direction of the intended receiver for acquiring the DPD coefficients.}

\begin{figure}[t!]
\centering
\includegraphics[width=1\linewidth]{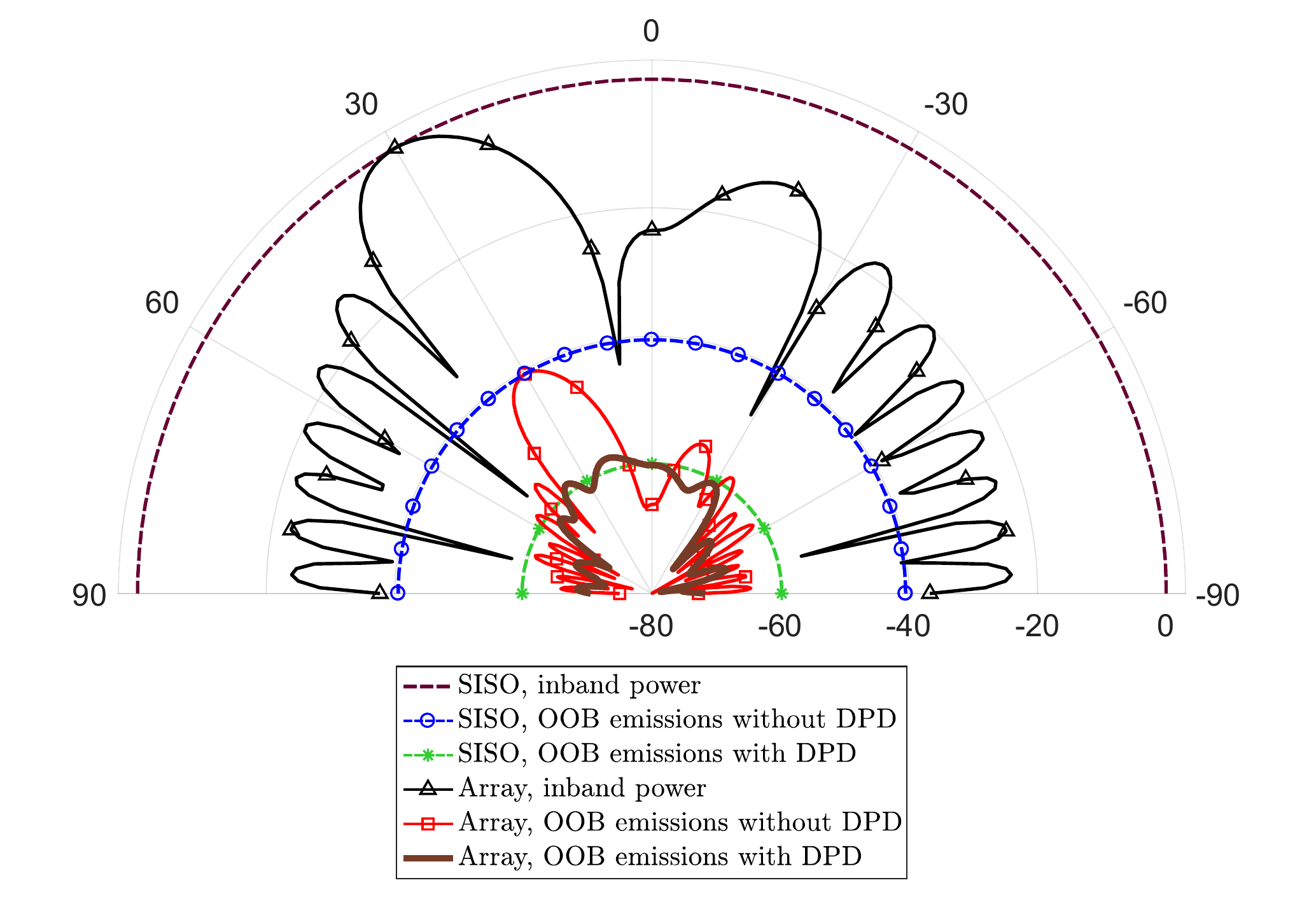}
\caption[]{In-band power and out-of-band emission patterns from a single antenna transmitter and from a 16-antenna array transmitter for all spatial directions ranging from $-90$ to $90$ degrees; r-axis represents relative powers, such that the received passband power at the intended RX direction, in both SISO and array cases, is normalized to $0$ dB. The in-band and OOB power levels are calculated over the allocated channel and the adjacent channel, respectively. The elements of the antenna array are uniformly distributed with a spacing of half the wavelength.}
\label{fig:Emissions_16ant}
\end{figure}

\textcolor{black}{
In Fig. \ref{fig:Emissions_16ant}, the inband power and OOB emission patterns in the spatial domain are shown for a single antenna transmitter, for reference, and for an array transmitter with sixteen antennas. In order to generate such patterns, it is necessary to take into account the individual antenna element radiation pattern, which is here assumed to be isotropic, and the array geometry, which we consider to be a uniform linear array with an antenna spacing of half the wavelength. The direction of the intended user is that of the direction of the main beam, which is 30 degrees in this numerical example.} 

\textcolor{black}{
The different power levels shown in the figure represent the total power for the inband and OOB emissions spanning the occupied bandwidth of the allocated channel and the adjacent channel, respectively, at different spatial directions. Since the received passband power is normalized to 0 dB, then taking this as the reference in-band power, the OOB patterns can be interpreted as the ACLR level in different spatial directions. For instance, the OOB emissions in the direction of the intended receiver (30 degrees) without predistortion have a level of $-40.48$ dB, while with predistortion it is $-61.42$ dB. These numbers correspond to ACLRs of $40.48$ dBc and $61.42$ dBc, respectively, as also indicated in Table \ref{tab:EVM_ACLR}. The corresponding ACLR numbers for another example direction of $-30$ degrees are $63.12$ dBc and $60.51$ dBc. Fig. \ref{fig:Emissions_16ant} thus constitutes a very useful and easily interpretable way to represent ACLR and its spatial characteristics in large array transmitters.} 

\begin{figure}[t!]
\centering
\includegraphics[width=1\linewidth]{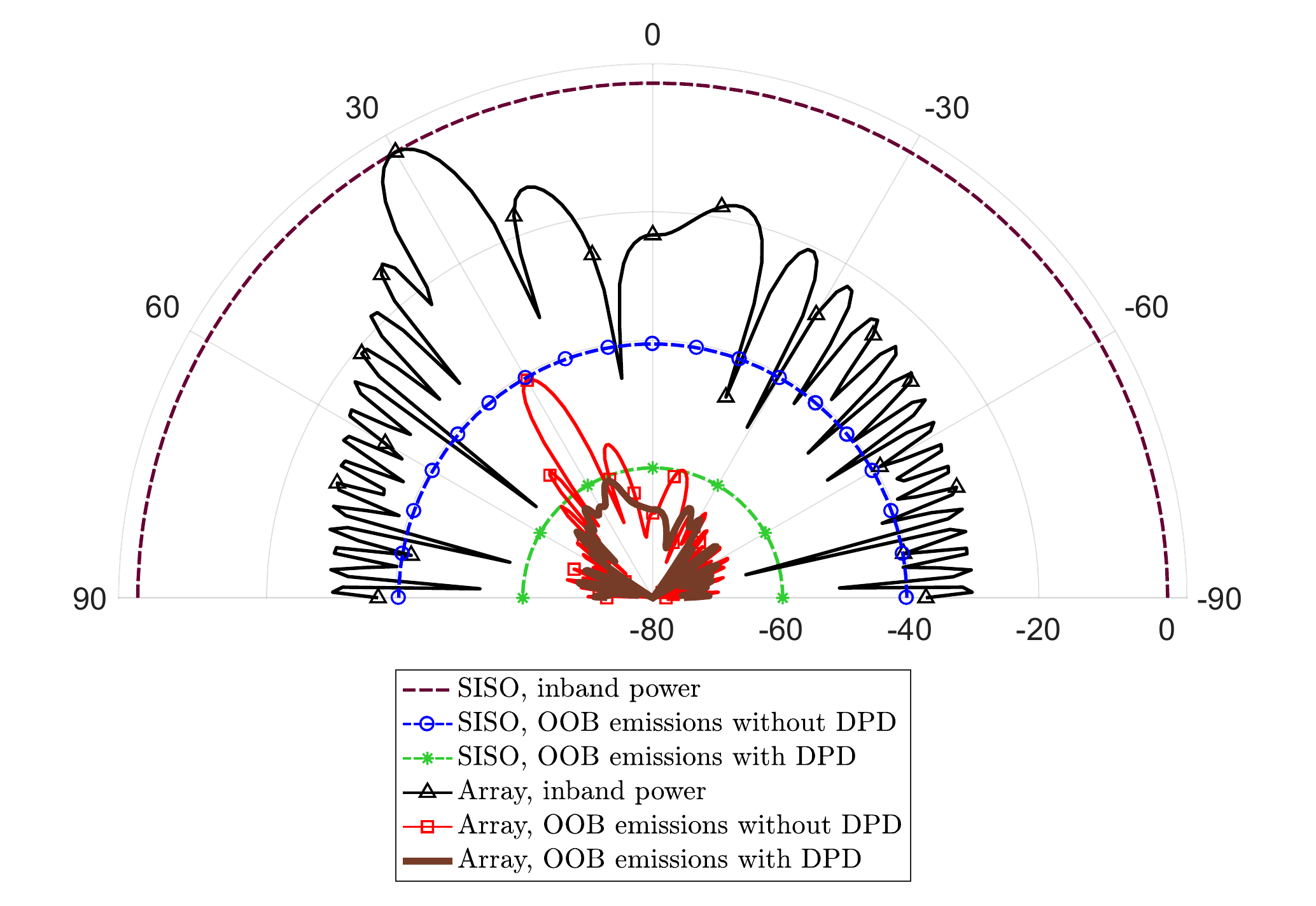}
\caption[]{In-band power and out-of-band emission patterns from a single antenna transmitter and from a 32-antenna array transmitter for all spatial directions ranging from $-90$ to $90$ degrees; r-axis represents relative powers, such that the received passband power at the intended RX direction, in both SISO and array cases, is normalized to $0$ dB. The in-band and OOB power levels are calculated over the allocated channel and the adjacent channel, respectively. The elements of the antenna array are uniformly distributed with a spacing of half the wavelength.}
\label{fig:Emissions_32ant}
\end{figure}

\textcolor{black}{When considering the inband and OOB emissions without DPD, the OOB emissions from the array are never larger than those of the single antenna case, as it was also concluded in \cite{OOB_Mollen}. This can be seen to be essentially true also after applying the proposed DPD. However, the OOB emissions in certain specific directions do exceed the reference single antenna case by a small margin (a few dBs at most), but are anyway kept at a sufficiently low level. This behavior is indeed due to the proposed algorithm primarily considering the emissions in the direction of the intended receiver, and the emissions in other spatial directions are defined by the joint effect of the DPD, the PA responses, and the antenna array beampattern. One can assume that the larger the antenna array and thus the beamforming gain are, the less probable it is for the array OOB emissions to exceed the reference single-antenna emissions. This is illustrated in Fig. \ref{fig:Emissions_32ant}, where a 32-antenna array is considered. Due to the higher spatial selectivity provided by the larger array, the OOB emissions are reduced such that they no longer exceed the single-antenna emissions in any spatial direction.}

\section{Conclusions}
\label{sec:Conclusions}
A novel reduced-complexity digital predistortion (DPD) solution was proposed in this paper for hybrid MIMO transmitters. 
The proposed DPD structure was developed taking into consideration the combined nonlinear effects of the PAs in a single sub-array of a hybrid MIMO transmitter.
The proposed DPD learning utilizes a combined feedback signal extracted from the PA units and thus requires only a single observation receiver chain. The proposed decorrelation-based learning aims at minimizing the correlation between the effective nonlinear distortion in the intended receiver direction, and specific nonlinear basis functions. Memory effects were considered in both the DPD structure and learning. The impact of amplitude and phase mismatches between the PA branches was also analyzed and shown to have a negligible effect under realistic assumptions. Practical simulations based on measured PA models were conducted to further demonstrate the effectiveness of the proposed solution. More than $10$ dB gain in ACLR was achieved when using the proposed DPD compared to using a single PA output for learning. \textcolor{black}{In addition, the spatial characteristics of the array out-of-band emissions with the proposed DPD structure were analyzed. While the largest reduction in the out-of-band emissions were shown to be available at the direction of the intended receiver, the emissions in the other spatial directions were also shown to be well-behaving and essentially at the same level or lower than those of the reference single-antenna transmitter, thanks to the combined effects of the DPD and beamforming. Thus, when it comes to evaluating traditional figures of merit, such as the ACLR, in antenna array transmitters, new approaches need to be considered since the out-of-band emissions behave differently than in single-antenna legacy systems.}

\bibliographystyle{IEEEbib}
\bibliography{Ref}

\begin{IEEEbiography}[{\includegraphics[width=1in,height=1.25in,clip,keepaspectratio]{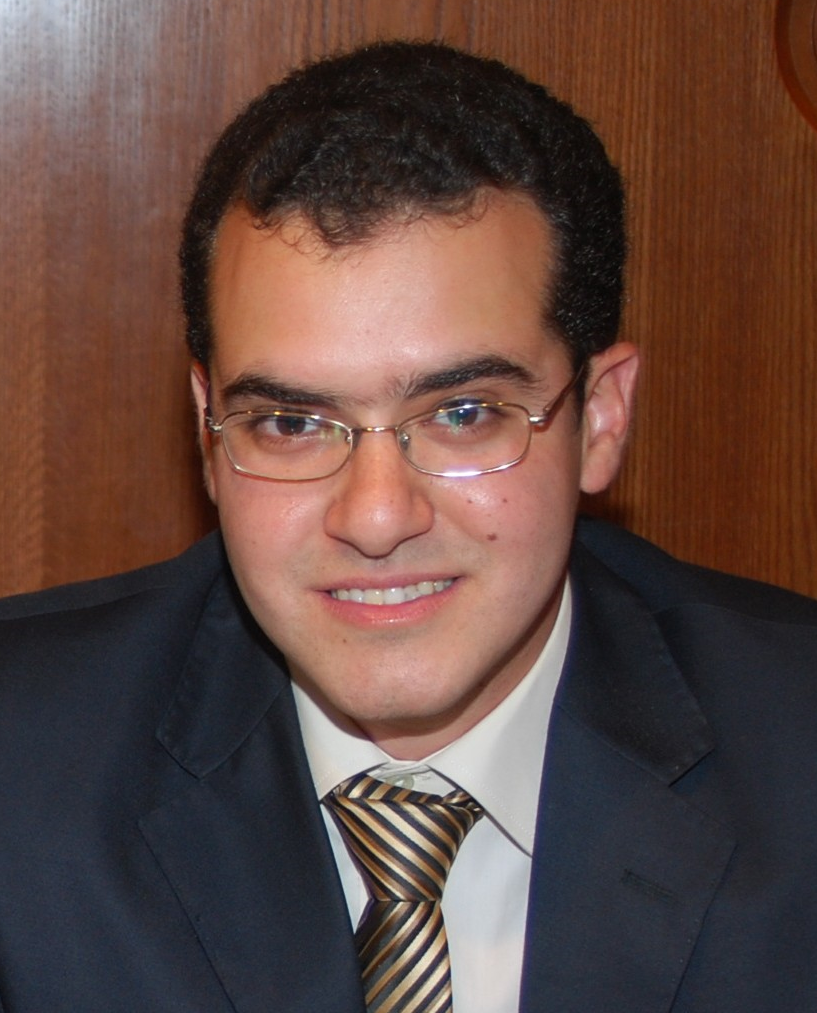}}]{Mahmoud Abdelaziz} received the D.Sc. (with honors) degree in Electronics and Communications Engineering from Tampere University of Technology, Finland, in 2017. He received the B.Sc. (with honors) and M.Sc. degrees in Electronics and Communications Engineering from Cairo University, Egypt, in 2006 and 2011, respectively. He currently works as a Postdoctoral researcher at Tampere University of Technology, Finland. Since February 2018, he has also been working with the electrical engineering department at the British University in Egypt. His research interests include statistical and adaptive signal processing in flexible radio transceivers.
\end{IEEEbiography}

\begin{IEEEbiography}[{\includegraphics[width=1in,height=1.25in,clip,keepaspectratio]{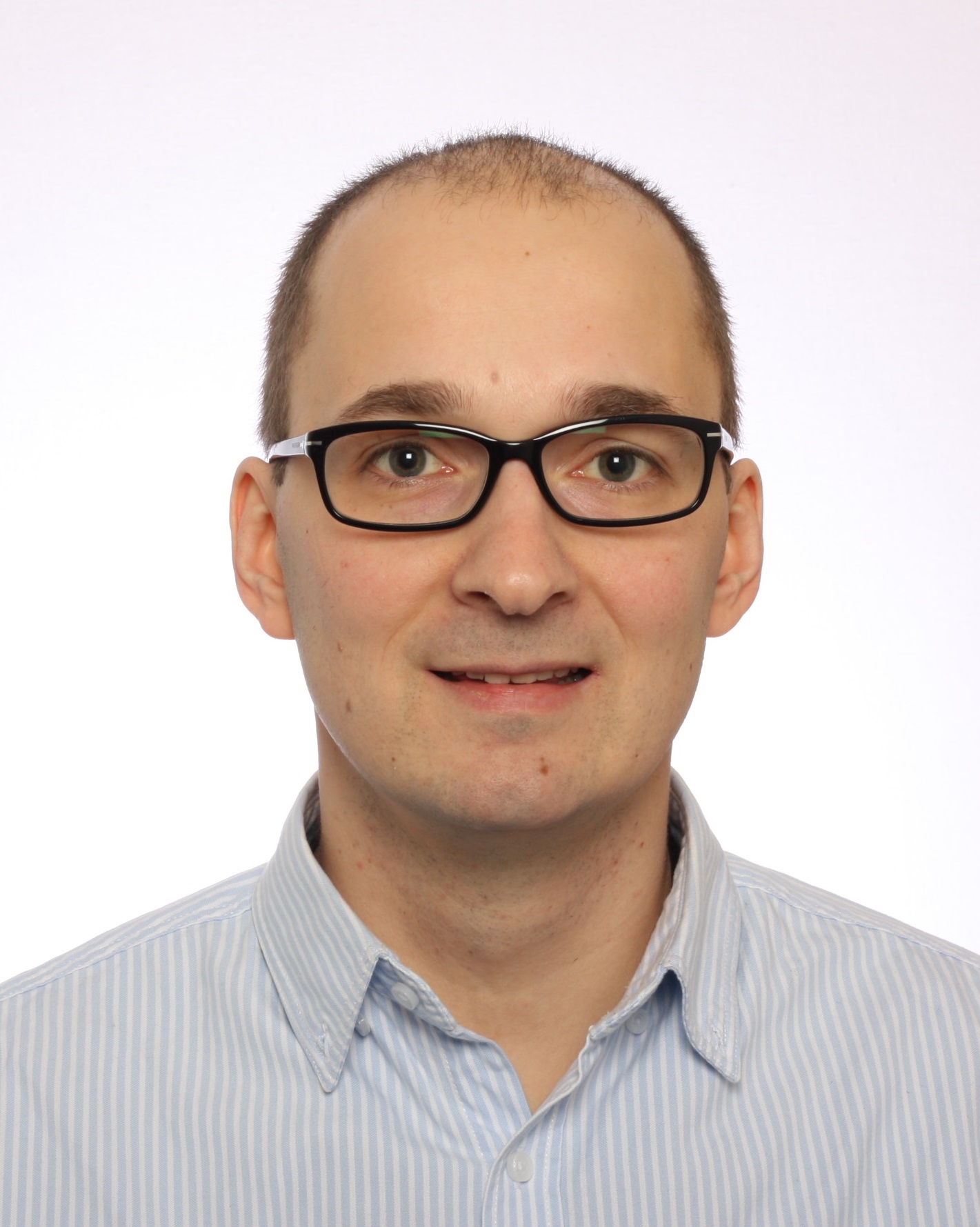}}]{Lauri Anttila} received the M.Sc. and D.Sc. (with honors) degrees in electrical engineering from Tampere University of Technology (TUT), Tampere, Finland, in 2004 and 2011. Since 2016, he has been a senior research fellow at the Laboratory of Electronics and Communications Engineering at TUT. In 2016-2017, he was a visiting research fellow at the Department of Electronics and Nanoengineering, Aalto University, Finland. His research interests are in signal processing for wireless communications, hardware constrained communications, and radio implementation challenges in 5G cellular radio, full-duplex radio, and large-scale antenna systems. 
\end{IEEEbiography}

\begin{IEEEbiography}[{\includegraphics[width=1in,height=1.25in,clip,keepaspectratio]{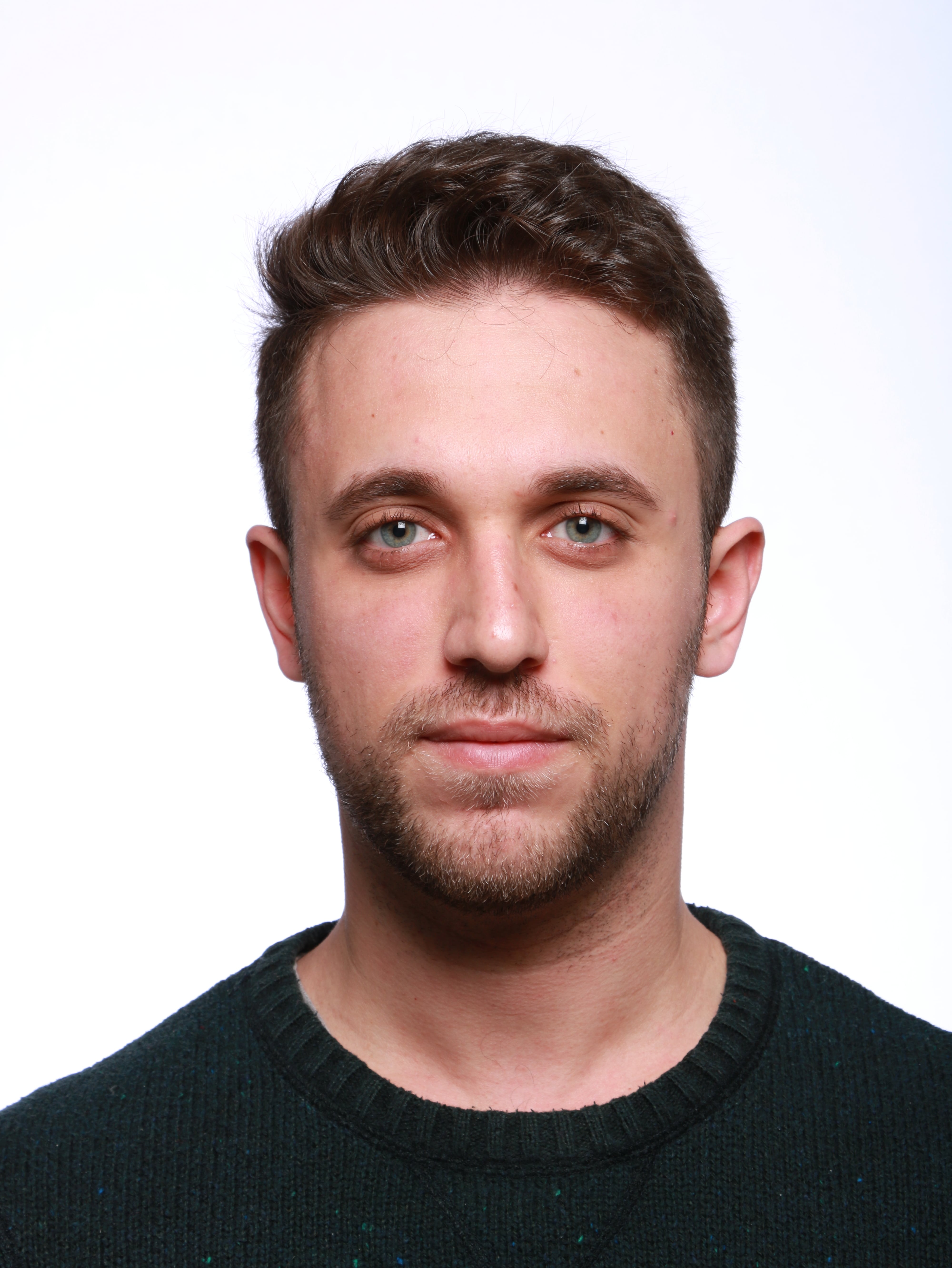}}]{Alberto Brihuega} received the B.Sc. and M.Sc. degrees in Telecommunications Engineering from Universidad Politecnica de Madrid, Spain, in 2015 and 2017, respectively. He is currently working towards the Ph.D. degree with Tampere University of Technology, Finland, where he is a researcher with the Laboratory of Electronics and Communications Engineering. His research interests include statistical and adaptive signal processing, as well as wideband digital predistortion and precoding techniques for massive MIMO.
\end{IEEEbiography}

\begin{IEEEbiography}[{\includegraphics[width=1in,height=1.25in,clip,keepaspectratio]{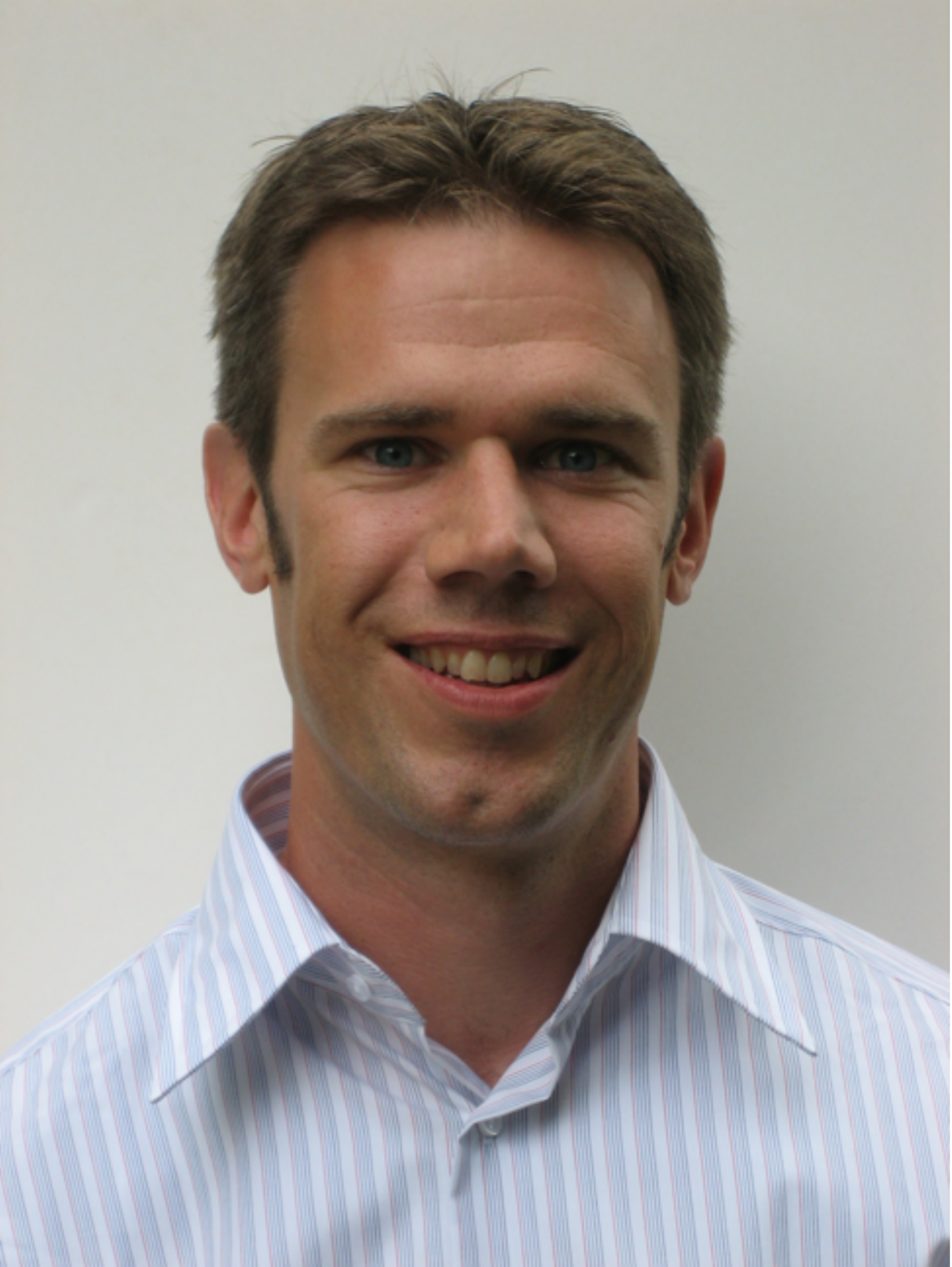}}]{Fredrik Tufvesson} received his Ph.D. in 2000 from Lund University in Sweden. After two years at a startup company, he joined the department of Electrical and Information Technology at Lund University, where he is now professor of radio systems. His main research interests is the interplay between the radio channel and the rest of the communication system with various applications in 5G systems such as massive MIMO, mm wave communication, vehicular communication and radio based positioning.
\end{IEEEbiography}

\begin{IEEEbiography}[{\includegraphics[width=1in,height=1.25in,clip,keepaspectratio]{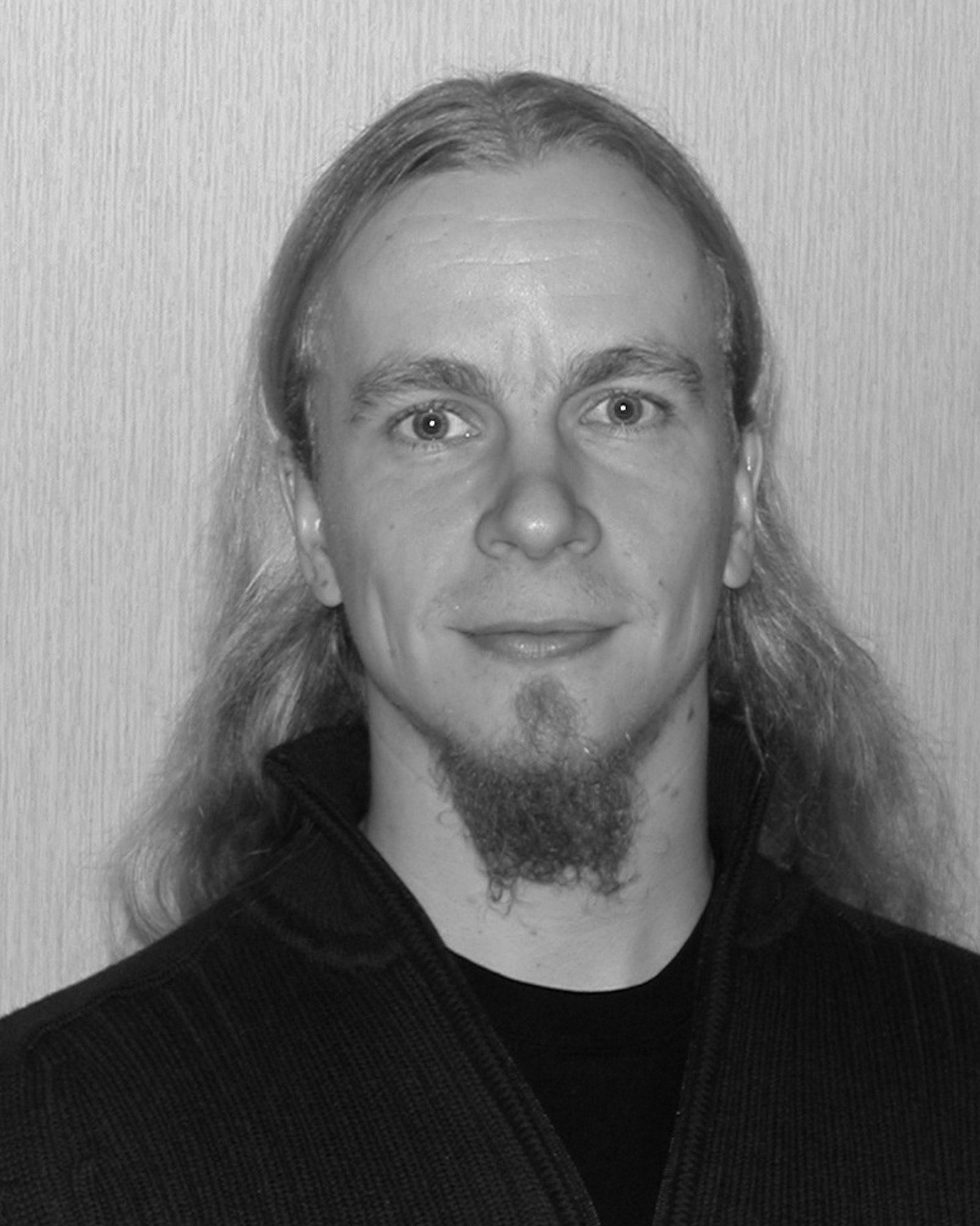}}]{Mikko Valkama} (S'00--M'01--SM'15) 
received the M.Sc. and Ph.D. degrees (both with honors) in electrical engineering (EE) from Tampere University of Technology (TUT), Finland, in 2000 and 2001, respectively. 
In 2003, he was working as a visiting post-doc research fellow with the Communications Systems and Signal Processing Institute at SDSU, San Diego, CA. Currently, he is a Full Professor and Laboratory Head at the Laboratory of Electronics and Communications Engineering at TUT, Finland. His general research interests include radio communications, radio signal processing and sensing, radio positioning, as well as 5G and beyond mobile radio networks. 
\end{IEEEbiography}

\end{document}